\documentclass[cameraready]{Interspeech}

\title{Geometric Second-Order Feature Correlation Learning for Self-Supervised Speech Emotion Recognition}

\author[affiliation={1}]{Shuanglin}{Li}
\author[affiliation={1}]{Ruxiao}{Qian}
\author[affiliation={2},correspondingauthor]{Siyang}{Song}

\address{
    $^1$ Xiangjiang Laboratory, Changsha, China
    $^2$ University of Exeter, Exeter, UK
}
\email{slay575@163.com, ruxiaoqian@gmail.com, s.song@exeter.ac.uk}


\usepackage{comment}
\usepackage{xurl}
\usepackage{bm}
\usepackage{hyperref}         
\usepackage{algorithm}
\usepackage{algpseudocode}
\usepackage{amsmath}  
\usepackage{amssymb}
\usepackage{graphicx} 
\usepackage{multirow}
\usepackage{siunitx} 
\usepackage{colortbl}

\newcommand{\second}[1]{%
    \classicobox{#1}%
}
\newcommand{\classicobox}[1]{%
    \colorbox{gray!30}{\hspace{-2pt}#1\hspace{-2pt}}%
}

\begin{document}

\maketitle

\begin{abstract}
Self-supervised learning (SSL) yields powerful, context-rich representations for speech emotion recognition (SER), yet the aggregation of these representations into holistic descriptors remains a bottleneck. Conventional first-order aggregation implicitly assumes feature independence, violating the latent Riemannian geometry and discarding higher-order relationships essential to the backbone’s representational power. To address this problem, this paper proposes a novel Second-Order Correlation (SOC) layer. Instead of treating features in isolation, our SOC models correlations among features as covariance descriptors to capture synergistic co-occurrence patterns, which act as discriminative signatures for robust emotion recognition. By mapping these descriptors from the Riemannian manifold to a Euclidean tangent space through Log-Euclidean mapping (LEM), our method preserves geometric integrity while enabling direct linear discriminative learning. Extensive experiments on ESD and RAVDESS datasets demonstrate that SOC recovers discriminative information lost in first-order pooling, effectively aggregating high-dimensional SSL features. 
\end{abstract}

\section{Introduction}

Deep self-supervised learning (SSL) has significantly advanced speech emotion recognition (SER) by providing powerful representation backbones. However, most downstream networks are predominantly based on simplistic first-order methods to aggregate the learned frame-level speech embeddings \cite{chen2023exploring, sharma2022multi,desplanques2020ecapa, pasad2021layer,leygue25_interspeech,su2020improving,stafylakis2023extracting,kakouros2023speech}. Such methods only treat latent components as uncorrelated entities and thus collapse the joint distribution of the feature space \cite{lin2015bilinear,li2017second}. 

This constraint is vital for SER, where emotional cues manifest through the synergistic interplay of prosodic and spectral dynamics \cite{niu2025learning}. Although such synergies are encoded in off-diagonal covariance elements, they are eliminated by linear operators—ranging from average pooling to attention mechanisms \cite{schuller2013computational, wagner2023dawn}. Consequently, these first-order strategies fail to capture the second-order dependencies that constitute distinctive emotional signatures.

\begin{figure*}[htbp]
    \centering
    \includegraphics[scale=0.23]{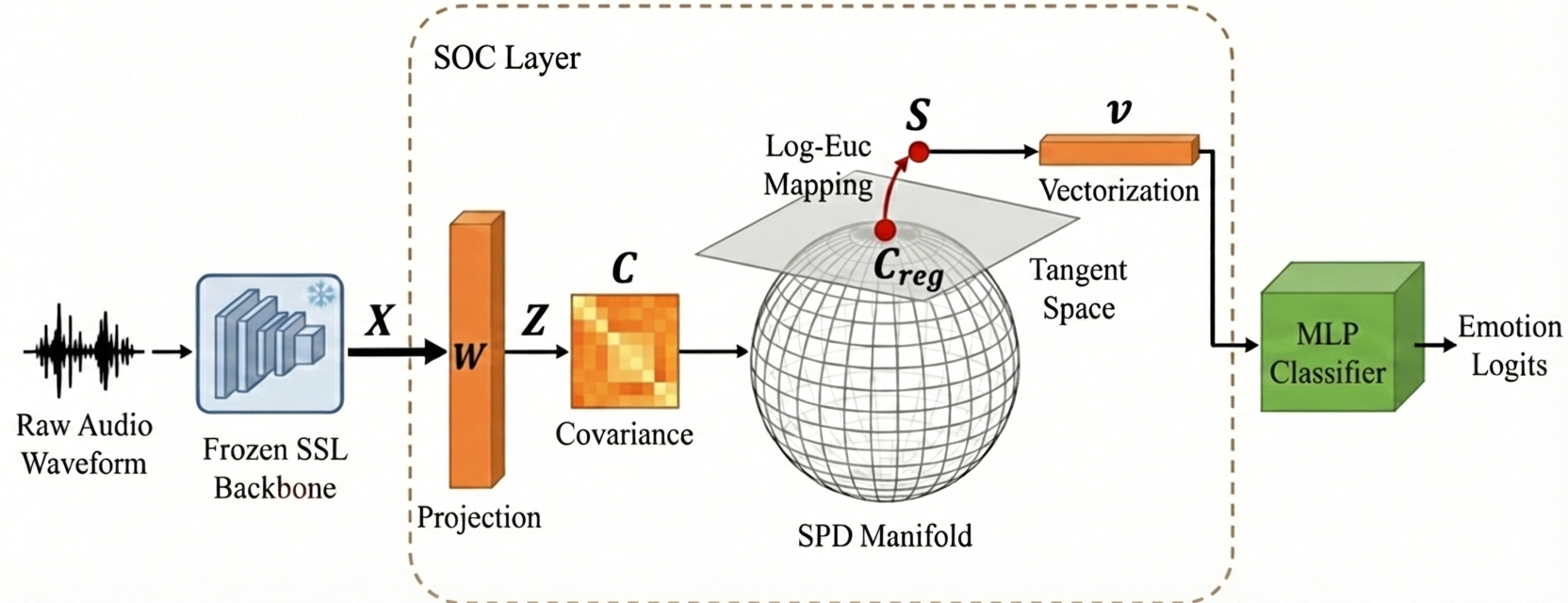}
    \caption{The overall framework of the proposed method. The raw audio is first processed by a frozen SSL backbone to extract high-level feature representations $\boldsymbol{X}$. These features are then mapped into a lower-dimensional space $\boldsymbol{Z}$ via a projection layer ($\boldsymbol{W}$). Within the SOC layer, the covariance matrices $\boldsymbol{C}$ is computed and regularized as $\boldsymbol{C}_{reg}$ to ensure it resides on the SPD manifold. Finally, the matrices are mapped to the tangent space $\boldsymbol{S}$ through Log-Euclidean mapping and vectorized into $\boldsymbol{v}$ for the downstream MLP classifier.}
    \label{fig:methods}
\end{figure*}

Early explorations \cite{tahir2013training, tuske2015integrating,ye2008speech} utilized polynomial expansions to capture these higher-order dependencies but were hindered by a prohibitive dimensionality explosion. When applied to high-dimensional SSL backbones (e.g., 768-D or 1024-D), the quadratic growth of parameters renders such methods computationally intractable (\textbf{Problem 1}). While contemporary techniques like global bilinear pooling \cite{lin2015bilinear} offer a compact alternative to mitigate this computational bottleneck, they fundamentally misinterpret the geometric nature of second-order statistics. By performing aggregation directly in Euclidean space without the requisite tangent space projection, these methods subject non-linear covariance descriptors to linear operations (e.g., averaging). Mathematically, such illegal operations force the resulting points to deviate from the manifold—a phenomenon known as the "swelling effect" \cite{arsigny2006log}. This deviation introduces spurious entropy: a form of geometric distortion that injects artificial variance into the descriptors, effectively blurring the decision boundaries between acoustically similar emotions (\textbf{Problem 2}). Robust SER requires efficient correlation modeling grounded in the manifold structure of SSL features, where eliminating spurious entropy preserves geometrically faithful and discriminative emotional signatures.

To bridge these gaps, this paper proposes a novel Second-Order Correlation (SOC) layer. SOC first projects high-dimensional SSL embeddings into a compact subspace via a learnable linear layer, enabling efficient and numerically stable covariance estimation. Subsequently, Log-Euclidean Mapping (LEM) is utilized to project these descriptors onto a tangent space, preserving their intrinsic geometric integrity while enabling standard Euclidean optimization. The most related work is HYFuse \cite{phukan25b_interspeech}, which attempts to leverage non-Euclidean geometry but relies on a preliminary vectorization step. Initial flattening discards structural correlations of the second-order manifold, reducing geometric operations to post-processing on compromised features. In contrast, SOC models correlations as covariance descriptors, using LEM to preserve geometric information lost in geometry-agnostic aggregation.

The main contributions of this work are three-fold: 1) We model SSL speech feature correlations as Symmetric Positive Definite (SPD) manifold-valued representations, leveraging non-Euclidean geometry to better characterize emotional ambiguity in SER.
2) We design SOC as a drop-in module that captures high-order statistics within a projected subspace, which effectively bypasses the dimensionality constraints and instability of high-dimensional SSL backbones; and 3) Extensive evaluations on ESD and RAVDESS datasets demonstrate that SOC consistently outperforms first-order baselines, validating the potential of manifold-based representations for the SER community. We provide our implementation at: \url{https://github.com/secret-code-source/SOC}.

\section{Proposed Method}

As illustrated in Fig. \ref{fig:methods}, our entire framework comprises three stages: (i) a frozen upstream SSL backbone for frame-level feature extraction; (ii) a novel SOC layer first projects features into a compact subspace to mitigate computational intractable, then models their correlations on the SPD manifold and maps them via LEM to preserve geometric integrity; and (iii) a standard MLP for the final speech emotion prediction. Throughout this paper, vectors are denoted as row vectors unless stated otherwise.

\begin{algorithm}[t]
\caption{Second-Order Correlation (SOC) Layer}
\label{alg:soc_pooling}
\begin{algorithmic}[1]
\renewcommand{\algorithmicrequire}{\textbf{Input:}}
\renewcommand{\algorithmicensure}{\textbf{Output:}}
\newcommand{\StepHeader}[1]{\vspace{0.15cm}\Statex \hspace*{-1.2em}\textbf{#1}}

\Require SSL features $\boldsymbol{X} \in \mathbb{R}^{T \times D_{in}}$, Subspace dimension $d$ 
\vspace{0.2em}
\Ensure Vectorized correlation features $\boldsymbol{v} \in \mathbb{R}^{d(d+1)/2}$

\StepHeader{Step 1: Input Centering} 
\State $\bar{\boldsymbol{x}} \leftarrow \frac{1}{T}\sum_{t=1}^{T}\boldsymbol{x}_t$ \hfill \Comment{Compute global temporal mean}

\StepHeader{Step 2: Subspace Projection}
\State $\boldsymbol{Z} \leftarrow (\boldsymbol{X} - \mathbf{1}\bar{\boldsymbol{x}}^\top) \boldsymbol{W}$ \hfill \Comment{Project centered features via $\boldsymbol{W}$}

\StepHeader{Step 3: Covariance Calculation \& Normalization}
\State $\boldsymbol{C} \leftarrow \frac{1}{T-1} \boldsymbol{Z}^\top \boldsymbol{Z}$ \hfill \Comment{Compute covariance}
\State $\hat{\boldsymbol{C}} \leftarrow \boldsymbol{C} / (\operatorname{tr}(\boldsymbol{C}) + \epsilon_{div})$ \hfill \Comment{Trace normalization}

\StepHeader{Step 4: Log-Euclidean Mapping}
\State $[\boldsymbol{U}, \boldsymbol{\Lambda}] \leftarrow \operatorname{EigDecomp}(\hat{\boldsymbol{C}} + \epsilon \boldsymbol{I}_d)$ \hfill \Comment{Eigen-Decomposition}
\State Let $\boldsymbol{\Lambda} = \operatorname{diag}(\lambda_1, \dots, \lambda_d)$ \hfill \Comment{Define eigenvalues}
\State $\boldsymbol{L} \leftarrow \operatorname{diag}(\log(\lambda_1), \dots, \log(\lambda_d))$ \Comment{Log-Eigenvalue}
\State $\boldsymbol{S} \leftarrow \boldsymbol{U} \boldsymbol{L} \boldsymbol{U}^\top$ \hfill \Comment{Map to the tangent space}

\StepHeader{Step 5: Half-Vectorization}
\State $\boldsymbol{v} \leftarrow \operatorname{vech}(\boldsymbol{S})$ \hfill \Comment{Extract lower triangular elements}
\State \Return $\boldsymbol{v}$

\end{algorithmic}
\end{algorithm}

\subsection{Upstream Feature Extraction}
We utilize frozen SSL backbones to extract final-layer and frame-level representations $\boldsymbol{X} \in \mathbb{R}^{T \times D_{in}}$ from raw speech. While this sequence captures rich contextual information, its high dimensionality poses a significant computational burden for subsequent second-order modelling, which typically scales quadratically with the feature size. To ensure a computationally tractable pipeline, $\boldsymbol{X}$ is passed to the proposed SOC layer for structured dimensionality reduction and geometric encoding.

\subsection{Second-Order Correlation (SOC) Layer}
\label{ssec:soc_pooling}

While standard aggregation schemes ignore feature correlations essential for emotional prosody, our SOC layer (Alg.~\ref{alg:soc_pooling}) explicitly captures these interactions. The module operates in two stages: (i) \textit{Subspace Projection and Manifold Construction}, which generates compact, scale-invariant descriptors; and (ii) \textit{Tangent Space Mapping}, which projects these Riemannian objects into a Euclidean space to bridge geometric incompatibilities with downstream networks.

\textbf{Subspace Projection and Manifold Construction} Directly computing covariance descriptors on high-dimensional SSL features $\boldsymbol{X} \in \mathbb{R}^{T \times D_\text{in}}$ introduces redundancy and computational instability \cite{tahir2013training, tuske2015integrating}. Therefore, we first seek a compact discriminative subspace by projecting $\boldsymbol{X}$ onto a lower-dimensional space through a learnable linear mapping $\boldsymbol{W} \in \mathbb{R}^{D_\text{in} \times d}$, where $d \ll D_\text{in}$. For each frame $t$, the centered feature vector $\boldsymbol{z}_t \in \mathbb{R}^{1 \times d}$ is obtained by factoring out the global temporal mean $\bar{\boldsymbol{x}}$ and applying the projection matrix $\boldsymbol{W}$ (\textbf{Steps 1 \& 2}):
\begin{equation}
\boldsymbol{z}_t = (\boldsymbol{x}_t - \bar{\boldsymbol{x}}) \boldsymbol{W}
\end{equation}
By stacking these centered vectors row-wise, the feature matrix $\boldsymbol{Z} \in \mathbb{R}^{T \times d}$ is obtained, which serves as the basis for capturing higher-order statistics. Subsequently, we compute the sample covariance matrix $\boldsymbol{C} \in \mathbb{R}^{d \times d}$ to capture the pairwise correlations between feature channels as:
\begin{equation}
\boldsymbol{C} = \frac{1}{T-1}\boldsymbol{Z}^\top \boldsymbol{Z}
\end{equation}
To achieve scale invariance and mitigate the influence of non-emotional energy variations, we apply Trace Normalization to derive the normalized descriptor $\hat{\boldsymbol{C}}$ (\textbf{Step 3}) as:
\begin{equation}
\hat{\boldsymbol{C}} = \frac{\boldsymbol{C}}{\operatorname{tr}(\boldsymbol{C}) + \epsilon_{div}}
\end{equation}
where $\operatorname{tr}(\boldsymbol{C})$ represents the total variation and $\epsilon_\text{div}$ is a small constant for numerical stability. This operation ensures that $\hat{\boldsymbol{C}}$ resides on a unit-trace SPD manifold, effectively decoupling the underlying correlation structure from absolute embedding magnitudes and providing a robust geometric basis for subsequent transformation.

\textbf{Tangent Space Mapping} Crucially, these SPD descriptors reside on a Riemannian manifold, creating a geometric incompatibility with Euclidean-based classifiers. 
To resolve this, we employ Log-Euclidean Mapping (LEM) to project the manifold onto a locally linear tangent space. By converting multiplicative geodesic distances into additive Euclidean distances, LEM flattens the non-linear geometry into a vector space where descriptors are treated as standard Euclidean tensors. Furthermore, to ensure the matrix logarithm is well-defined, we incorporate a minor perturbation $\epsilon \boldsymbol{I}_d$ during the eigen-decomposition of $\hat{\boldsymbol{C}}$ (Step 4):
\begin{equation}
\hat{\boldsymbol{C}} + \epsilon \boldsymbol{I}_d = \boldsymbol{U} \boldsymbol{\Lambda} \boldsymbol{U}^\top
\end{equation}
where $\boldsymbol{U}$ is an orthogonal matrix of eigenvectors and $\boldsymbol{\Lambda} = \operatorname{diag}(\lambda_1, \dots, \lambda_d)$ contains the strictly positive eigenvalues. The matrix logarithm is then computed in the spectral domain to obtain the log-mapped eigenvalue matrix as:
\begin{equation}
\boldsymbol{L} = \operatorname{diag}(\log(\lambda_1), \dots, \log(\lambda_d))
\end{equation}
where the $\log(\cdot)$ operator transforms the non-linear Riemannian geometry into a flat tangent space by mapping multiplicative geodesic distances into additive Euclidean distances.
Subsequently, the manifold-valued descriptor is projected onto the tangent space via the following transformation:
\begin{equation}
\boldsymbol{S} = \boldsymbol{U} \boldsymbol{L} \boldsymbol{U}^\top
\end{equation}
The reconstructed matrix $\boldsymbol{S} \in \mathbb{R}^{d \times d}$ serves as a linearized Euclidean proxy of the original Riemannian descriptor. To eliminate numerical redundancy, we apply half-vectorization (Step 5) to isolate its unique statistical profile:
\begin{equation}
\boldsymbol{v} = \operatorname{vech}(\boldsymbol{S}) \in \mathbb{R}^{\frac{d(d+1)}{2}}
\end{equation}
By extracting the $\frac{d(d+1)}{2}$ unique elements from the symmetric matrix $\boldsymbol{S}$, the $\operatorname{vech}(\cdot)$ operator yields a minimally sufficient representation. This preserves the full second-order profile in a compact vector format, ensuring compatibility with standard downstream classifiers while eliminating the numerical redundancy of duplicate symmetric entries.

\subsection{Downstream Classification}
\label{ssec:classification}
As the final stage, an MLP maps the vectorized features $\boldsymbol{v}$ to emotion logits for classification. The entire framework is optimized end-to-end. Specifically, the SOC layer's reliance on differentiable matrix operations—including eigen-decomposition—ensures seamless gradient flow, enabling joint optimization of the subspace projection and the classifier to maximize discriminative power.

\section{Experiments and Results}

\subsection{Datasets}
We evaluate our proposed method on two widely used emotional speech datasets: ESD \cite{zhou2021esd} and RAVDESS \cite{livingstone2018ryerson}.

\begin{itemize}
     \item \textbf{ESD:} is a bilingual corpus (English/Mandarin) containing 35,000 utterances from 20 speakers (10 per language; 5 per gender). Each speaker provides 350 parallel utterances across five emotions (Neutral, Happy, Sad, Angry, and Surprise), recorded at 16 kHz with an average duration of 2.5 seconds.
     \item \textbf{RAVDESS:} We utilize the audio-only subset, comprising 1,440 recordings from 24 professional actors (12 per gender). It covers eight emotions: Neutral (96 samples) and seven others (Calm, Happy, Sad, Angry, Fearful, Disgust, Surprise) at 192 samples each.
\end{itemize}

\noindent We follow the standardized preprocessing and speaker-independent evaluation protocol defined in EmoBox \cite{ma2024emobox}. All audio signals are resampled to 16~kHz and mono-converted. Further details regarding partitioning configurations and hyper-parameters are available in \cite{ma2024emobox}.

\subsection{Implementation Details}

\noindent \textbf{Training Configurations:} Models are implemented in PyTorch and trained on a NVIDIA RTX 4090 for 100 epochs using Cross-Entropy loss. We use the AdamW \cite{loshchilov2019decoupled}  optimizer (weight decay: $10^{-4}$) with a linear scheduler (10\% warmup; peak LR: $1 \times 10^{-4}$). Batch sizes are 64 for ESD and 32 for RAVDESS. \\
\noindent \textbf{Feature Extraction:} We utilize three frozen SSL backbones: Wav2Vec 2.0 \cite{baevski2020wav2vec}, HuBERT \cite{hsu2021hubert}, and WavLM \cite{chen2022wavlm} (all base versions) to extract frame-level features ($D_{in} = 768$). \\
\noindent \textbf{Evaluation:} Following the speaker-independent protocol in EmoBox \cite{ma2024emobox}, we evaluate performance using Weighted Accuracy (WA), Unweighted Accuracy (UA), and Macro F1-score. We employ $k$-fold cross-validation partitioned strictly by speaker groups, specifically, 5-fold for ESD and 6-fold for RAVDESS, with 20\% of the training data in each fold reserved for validation. \\
\noindent \textbf{Baselines:} We compare the SOC layer against three representative methods: GAP \cite{ma2024emobox}, which uses the global feature mean as a foundational baseline; ASP \cite{okabe18_interspeech}, which adds channel-wise standard deviation to capture marginal variability; and FA \cite{kim22d_interspeech}, an amplitude-aware aggregation mechanism that learns non-uniform weighting across frames to prioritize informative feature representations.

\begin{table*}[t]
    \centering
    \small 
    \renewcommand{\arraystretch}{1.2} 
    
    \sisetup{
        detect-weight=true, 
        detect-inline-weight=math,
        table-format=2.2,
        table-parse-only 
    }
    
    \caption{Performance comparison on ESD and RAVDESS datasets using three frozen SSL backbones. All values are reported in percentage (\%). The best results are \textbf{bolded}, and the second best are \colorbox{gray!20}{shaded}. ($\uparrow$ indicates higher is better.)}
    \label{tab:main_results}
    
    \begin{tabular}{@{} l l *{6}{S} @{}}
        \toprule
        \multirow{2.5}{*}{\textbf{Backbone}} & \multirow{2.5}{*}{\textbf{Method}} & \multicolumn{3}{c}{\textbf{ESD (5-fold) $\uparrow$}} & \multicolumn{3}{c}{\textbf{RAVDESS (6-fold) $\uparrow$}} \\
        \cmidrule(lr){3-5} \cmidrule(l){6-8} 
        & & {\textbf{WA}} & {\textbf{UA}} & {\textbf{F1}} & {\textbf{WA}} & {\textbf{UA}} & {\textbf{F1}} \\
        \midrule
        
        \multirow{5}{*}{Wav2Vec 2.0} 
        & GAP           & 67.18 & 67.18 & 66.75 & 54.25 & 54.22 & 53.64 \\
        & ASP           & 63.83 & 63.83 & 63.18 & 52.50 & 54.30 & 53.68 \\
        & FA            & {\second{68.94}} & {\second{68.94}} & {\second{68.42}} & {\second{56.27}} & {\second{56.45}} & 55.12 \\
        \cmidrule(lr){2-8}
        & SOC (w/o LEM) & 68.30 & 68.30 & 67.95 & 55.42 & 55.80 & {\second{55.71}} \\
        & \textbf{SOC (Ours)} & \textbf{71.86} & \textbf{71.86} & \textbf{71.23} & \textbf{58.67} & \textbf{58.52} & \textbf{57.96} \\
        
        \addlinespace[0.5em]
        \midrule 
        \addlinespace[0.5em]
        
        \multirow{5}{*}{HuBERT} 
        & GAP           & 71.38 & 71.38 & 71.19 & 65.24 & 64.92 & 64.94 \\
        & ASP           & 65.32 & 65.32 & 64.96 & 62.50 & 63.28 & 62.69 \\
        & FA            & {\second{72.48}} & {\second{72.48}} & {\second{72.10}} & 66.92 & 66.02 & 66.46 \\
        \cmidrule(lr){2-8}
        & SOC (w/o LEM) & 72.05 & 72.05 & 71.84 & {\second{68.10}} & {\second{67.85}} & {\second{67.52}} \\
        & \textbf{SOC (Ours)} & \textbf{73.50} & \textbf{73.50} & \textbf{72.82} & \textbf{69.75} & \textbf{69.38} & \textbf{69.61} \\
        
        \addlinespace[0.5em]
        \midrule 
        \addlinespace[0.5em]
        
        \multirow{5}{*}{WavLM} 
        & GAP           & 69.49 & 69.49 & 69.21 & 60.83 & 61.33 & 60.95 \\
        & ASP           & 66.71 & 66.71 & 66.78 & 63.45 & 63.83 & 63.43 \\
        & FA            & {\second{71.12}} & {\second{71.12}} & {\second{70.85}} & 66.25 & 66.89 & 67.60 \\
        \cmidrule(lr){2-8}
        & SOC (w/o LEM) & 70.82 & 70.82 & 70.15 & {\second{67.30}} & {\second{69.42}} & {\second{68.95}} \\
        & \textbf{SOC (Ours)} & \textbf{72.61} & \textbf{72.61} & \textbf{71.48} & \textbf{68.74} & \textbf{71.35} & \textbf{70.87} \\
        \bottomrule
    \end{tabular}
\end{table*}

\subsection{Results and Analysis}
Table 1 summarizes the performance on ESD and RAVDESS across three frozen SSL backbones. Our proposed SOC consistently outperforms all baselines, demonstrating robust universality. Notably, on the Wav2Vec 2.0 backbone, SOC exceeds the standard GAP by 4.68\% on ESD and 4.42\% on RAVDESS. This superiority persists across HuBERT and WavLM, with SOC achieving a peak WA of 73.50\% on ESD. Furthermore, SOC exhibits exceptional stability in data-scarce scenarios like RAVDESS; while ASP occasionally falters, SOC maintains a significant lead—outperforming the strongest baseline (FA) by 2.49\% on WavLM. These results validate that SOC's manifold-based aggregation more effectively captures discriminative high-order dependencies than traditional first-order or marginal statistics.

\subsection{Ablation Study}
To verify the core components of the SOC layer, we analyze the necessity of LEM and the sensitivity of subspace dimension $d$, with ablation results for the former reported in Table 1 and Figure 2.

Table 1 presents the performance of the variant excluding LEM (SOC w/o LEM). Removing this projection causes a consistent performance drop, notably 1.45\% on ESD and 1.65\% on RAVDESS for the HuBERT backbone. As raw covariance features inhabit a non-Euclidean manifold; without tangent space mapping, this metric mismatch distorts class boundaries and impairs linear classification. LEM rectifies the distribution by aligning geodesic distances with Euclidean space, "untangling" the features from the manifold's curvature to achieve linear separability.

The subspace dimension $d$ balances structural richness and statistical stability. As shown in Fig. 2, performance follows a unimodal trend: low dimensions cause correlation starvation, limiting the capture of complex SSL channel interactions. Excessive dimensions, on the other hand, trigger the curse of dimensionality, where quadratic growth of parameters ($d(d+1)/2$) leads to singular covariance calculation. This instability causes eigenvalue dispersion and spectrum noise, distorting the LEM and Riemannian manifold's geometry. Thus, the optimal $d$ must balance representational capacity with noise redundancy to maintain a stable, discriminative tangent space.


\begin{figure}
    \centering
    \includegraphics[scale=0.50]{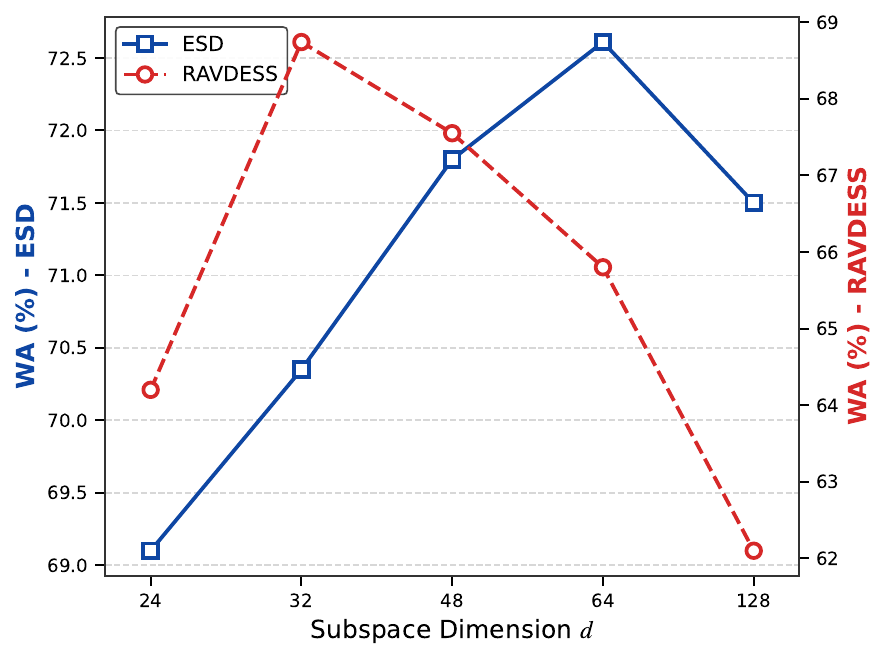}
    \caption{Impact of the subspace dimension $d$ on the performance of ESD and RAVDESS datasets. The left and right y-axes represent the WA for ESD and RAVDESS, respectively.}
\end{figure}

\subsection{Feature Distribution Visualization}
To qualitatively assess the learned representations, we visualize the feature space using t-SNE as shown in Fig. 3. Whereas GAP suffers from semantic fragmentation, SOC employs geometric correlations to unify Surprise and Sad samples into cohesive clusters. Furthermore, SOC addresses the ambiguity observed in GAP—particularly between Angry and Neutral classes by utilizing a Riemannian manifold to contract the former into a dense core and drive the latter to the periphery, successfully inducing a clear discriminative margin.

\begin{figure}
    \centering
    \includegraphics[scale=0.20]{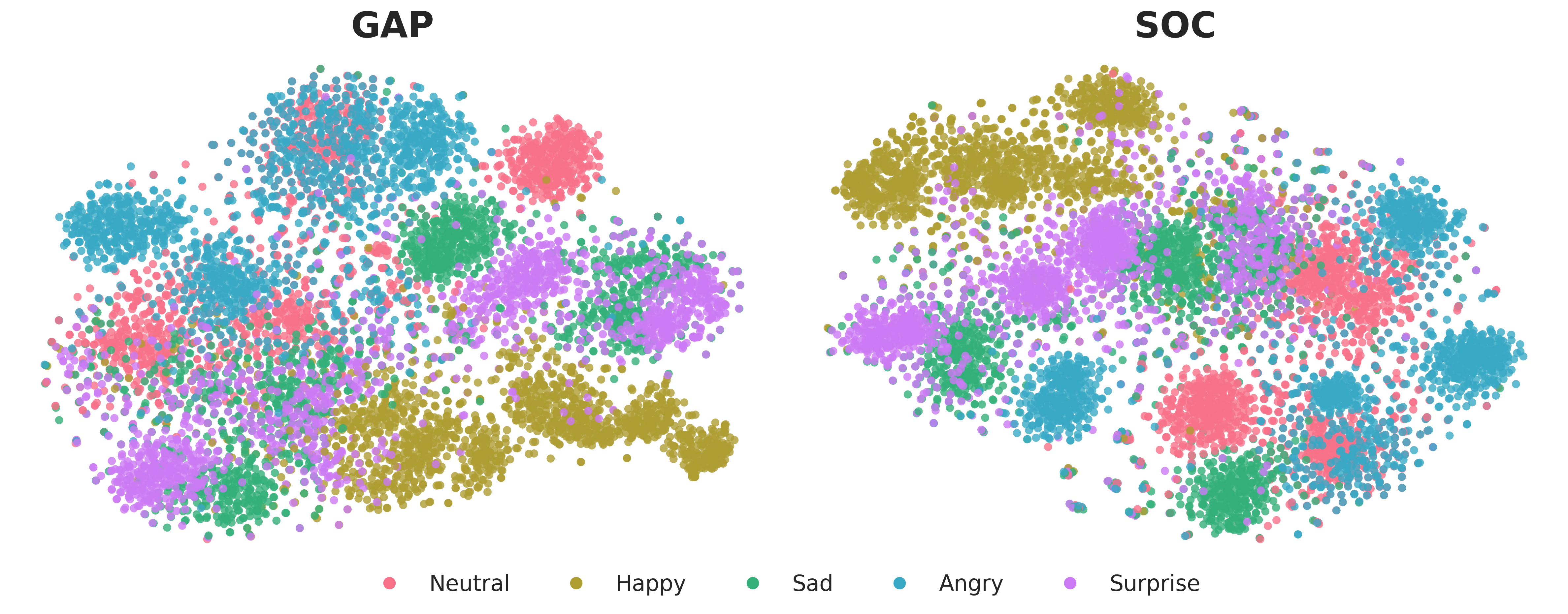}
    \caption{t-SNE visualization of WavLM feature distributions on ESD. Left: GAP ; Right: SOC}
\end{figure}

\section{Conclusion and Future Work}
In this paper, we presented the SOC layer, which captures joint feature dependencies by mapping covariance descriptors to a tangent space via LEM. By preserving geometric integrity, SOC consistently outperforms first-order baselines, demonstrating that geometric awareness is crucial for robust SER. 
\newpage

\bibliographystyle{IEEEtran}
\bibliography{mybib}

\end{document}